\documentclass[a4paper,fleqn,usenatbib]{mnras}
\usepackage[T1]{fontenc}
\usepackage{ae,aecompl}
\usepackage{graphicx}
\usepackage{amssymb}
\usepackage{amsmath}
\usepackage{mathptmx}
\usepackage{epstopdf}
\usepackage{booktabs}
\usepackage{float}
\usepackage{epsfig,color,ulem}
\usepackage{tabularx}
\newcommand{\A}{{\it A }}
\newcommand{\B}{{\it B }}
\newcommand{\C}{{\it C }}

\newcommand{\msun}{\,{\rm M_{\odot}}}
	
	\title[The cocoon emission - an EM counterpart to GW]{The cocoon emission - an electromagnetic counterpart to gravitational waves from neutron star mergers}
	
	\author[Gottlieb, Nakar \& Piran]{
		Ore Gottlieb,$^{1}$\thanks{oregottlieb@mail.tau.ac.il}
		Ehud Nakar,$^{1}$
		Tsvi Piran$^{2}$
		\\
		$^{1}${The Raymond and Beverly Sackler School of Physics and
			Astronomy, Tel Aviv University, Tel Aviv 69978, Israel}\\
		$^{2}${Racah Institute of Physics, The Hebrew University of
			Jerusalem, Jerusalem 91904, Israel}
	}
	\pubyear{2017}
\begin{document}
	\label{firstpage}
	\pagerange{\pageref{firstpage}--\pageref{lastpage}}
	\maketitle	
\begin{abstract}
Short Gamma-Ray Bursts (SGRBs) are believed to arise from compact binary mergers (either neutron star-neutron star or black hole-neutron star). If so their jets must penetrate outflows that are ejected during the merger. As a jet crosses the ejecta it dissipates its energy, producing a hot cocoon which surrounds it. We present here 3D numerical simulations of jet propagation in mergers' outflows and we calculate the resulting emission. This emission consists of two components: the cooling emission, the leakage of the thermal energy of the hot cocoon, and the cocoon macronova that arises from the radioactive decay of the cocoon's material. This emission gives a brief ($\sim$ one hour) blue, wide angle signal. While the parameters of the outflow and jet are uncertain, for the configurations we have considered the signal is bright ($\sim$ -14 -- -15 absolute magnitude) and outshines all other predicted UV-optical signals. The signal is brighter when the jet breakout time is longer and its peak brightness does not depend strongly on the highly uncertain opacity. A rapid search for such a signal is a promising strategy to detect an electromagnetic merger counterpart. A detected candidate could be then followed by deep IR searches for the longer but weaker macronova arising from the rest of the ejecta. 
\end{abstract}
\begin{keywords}
		{gamma-ray burst: short | stars: neutron | gravitational waves | methods: numerical}
\end{keywords}
\section{Introduction}
\label{introduction}	
Binary neutron star (NS) mergers are leading candidate sources for detection of gravitational wave signals by Advanced LIGO, Virgo and KAGRA. The localization of the gravitational radiation signal is rather poor and an electromagnetic (EM) counterpart is essential to identify the host and extract much more information on the merger process \citep{Kochanek1993}. These mergers are most likely accompanied by short gamma-ray bursts (SGRBs) \citep{Eichler1989}. However, these bursts are beamed and in most cases we won't observe the accompanying GRB (see \citealt{nakar2007} for a review).

Numerical simulations \citep[e.g.][]{Rosswog1998,Hotokezaka2011,Fernandez2013,Hotokezaka2013,Bauswein2013,Rosswog2013b,Perego2014,Siegel2014,Wanajo2014,Just2015} suggest that the mergers are accompanied by significant mass ejection from several sources \citep[see e.g.][for a review]{Hotokezaka2015,Metzger2016}. The ejected matter is enriched by heavy unstable nuclei whose radioactive decay power a macronova (also called kilonova) \citep{Li1998,Kulkarni2005,Metzger2010}. The interaction of the expanding ejecta with the surrounding medium produces, at a later stage, a radio flare lasting months to years \citep{Nakar2011}. Here we discuss yet another electromagnetic signal. The one that arises from the hot cocoon that is generated by the interaction between the SGRB jet and the ejecta.

If the SGRB jet is launched after at least some of the ejecta have expanded, then its propagation through the ejecta inflates a hot cocoon (\citealt{
Murguia-Berthier2014,Nagakura2014,Lazzati2016,Murguia-Berthier2017} and \citealt{Nakar2017}, hereafter NP17). As the jet interacts with the ejecta it forms a double shock structure at its head \citep{Matzner2003,Bromberg2011}. The reverse shock that crosses the jet material heats it up and divert it into an inner cocoon surrounding the jet. The ejecta that crosses the forward shock forms the outer cocoon. As long as the jet is within the ejecta it deposits most of its energy into the cocoon.

Once the cocoon breaks out of the ejecta it spreads sideways and radiates. This emission is spread over a wide opening angle making it an excellent potential EM counterpart to the gravitational waves source. 
This emission resembles to some extent a supernova emission. At first the hot cocoon expands and cools adiabatically while some radiation leaks out from its photosphere. {At the same time the cocoon, like the rest of the ejecta, is heated up by the radioactive decay of the r-process elements that were synthesized at the base of the wind.} If the inner cocoon (produced by the shocked jet material) is relativistic, its interaction with the external medium may also produce later an afterglow that is observed over a wider angles than the GRB's afterglow. 
{NP17 and \cite{Lazzati2016} discussed analytically the cocoon emission from SGRBs. NP17 provided a general framework to calculate all three components (i.e., cooling, radioactive and afterglow) based on the properties of the cocoon upon breakout. They pointed out that analytical model faces difficulties in modeling the mildly-relativistic cocoon material and the amount of mixing between the inner and outer cocoons, which is critical for assessing some of the emission components.} Here we use numerical simulations to first explore the jet and cocoon propagation within the ejecta and then to calculate the cocoon's signature. 

The time it takes the jet to drill through the ejecta, and in turn the energy of the cocoon, depends on the specific chain of events that take place between the final stages of the merger and the launch of the jet. A plausible scenario is one in which following the merger a hypermassive neutron star is formed \citep{Sekiguchi2011,Hotokezaka2012}, alongside a non-negligible ejected mass of $ M_{\rm{ej}} \approx 10^{-2}\msun $ \citep{Rosswog1998,Hotokezaka2013}. Shortly after the merger, the hypermassive neutron star collapses to a black hole and a jet is launched. In this model the jet is launched with a time delay relative to the merger, which is typically assumed to be a fraction of a second.
During this time the ejecta has propagated outward and the jet is launched into an extended dense medium. The time delay between the collapse and the launch of the jet, as well as the time it takes the jet to break out of the ejecta, is not well constrained by theoretical considerations. However, indirect observational evidence do provide us with some clues on the time it takes the jet to break out of the ejecta. \cite{Moharana2017} have shown that the distribution of SGRB durations suggests that the jet is launched for at least a few hundred millisecond in order for it to break out of the ejecta. In such a case the cocoon carries an energy that is comparable to that of the SGRB itself and the cocoon breakout radius is $\gtrsim 10^9$ cm.

For this paper we use the canonical ejecta model used by \citet{Nagakura2014} that derive an analytic wind model based on the numerical simulations of \citet{Hotokezaka2012}. While this model was derived based on simulations that found that the immediate outcome of the merger is a hyper-massive neutron star, our calculations do not depend necessarily on this specific scenario, and they are applicable to a range of merger models that involve a jet that propagates a significant fraction of a second within a quasi-spherical outflow.

The outline of this paper is as follows. The numerical setup for the hydrodynamic simulations is presented in $ \S $ \ref{sec_numerical} and their results are discussed in $ \S $ \ref{sec_hydro}. In $ \S $ \ref{sec_emission} we derive the cocoon emission that arises from the results of the hydrodynamic simulations. We first describe the
numerical method we use to calculate the emission, we then present the resulting light curves and discuss their observational implications. We conclude and summarize in $ \S $ \ref{sec_summary}. 
		
\section{Numerical Setup}
\label{sec_numerical}
Our simulation contains two components, a Newtonian spherically isotropic ejecta and a narrowly collimated ultra-relativistic jet. The spherical wind is set as initial conditions following the model of \citet{Nagakura2014} where $t=0$ corresponds to the merger time. At this time the wind ranges between $r_{esc} = 4.5 \times 10^7~\rm{cm} $ and $r_{max}=1.3 \times 10^8~\rm{cm}$. The wind expands homologously, namely,
\begin{equation} \label{v_wind}
	v(r) = v_{max}\bigg(\frac{r}{r_{max}(t)}\bigg)~,
\end{equation}
where $ v_{max} $ is the velocity of the outmost part of the ejecta and $r_{max}(t)$ is its radius at time $t$. The density profile satisfies $ \rho \propto r^{-3.5}~ $ and the initial pressure follows $ p \propto \rho^{4/3} $. The mass-weighted average velocity of the wind is $v_{avg} \approx v_{max}/2$.
The density normalization is such that the total (two-sided) ejecta mass at $t=0$ is $ 10^{-2}\msun $. During the simulation we continue to inject a homologous spherical wind at the boundary, reducing its mass flux continuously so the ejecta mass is dominated by the mass at $t=0$

The relativistic jet is injected along the $z$-axis as a boundary condition at $z_{base}=r_{esc}$. Its injection starts at $t_{inj}$, which is the delay between the merger and the beginning of the jet launching. Once the the jet injection starts it continues steadily throughout the simulation duration. We use the injection method of \cite{Mizuta:2013}, where the jet is injected through a cylindrical nozzle with an initial Lorentz factor $ \Gamma_{j,0}$ and a rest frame specific enthalpy $h_{j,0} \gg 1$. The jet then accelerates and spreads quickly to an opening angle $\theta_j=1/f \Gamma_{j,0}$, where $f=1.4$ \citep{harrison17}, and a terminal Lorentz factor $ \Gamma_{j,0} h_{j,0}$. We use a nozzle with a radius $ r_{j,0} = 10^7~\rm{cm}$ and a homogeneous distribution across its planar cross section. We take $h_{j,0}=20$, and an initial Lorentz factor of $ \Gamma_{j,0} = 4$, the corresponding jet opening angle is $\theta_j = 10^\circ $ and the jet terminal Lorentz factor is $80$.
The jet total luminosity is set to $ L_j = 4\times10^{50}~\rm{erg~s^{-1}} $, so in our simulation which covers only one side of the jet we inject half of this luminosity. With these parameters the average jet's head velocity during its propagation in the ejecta is $\sim 0.5 c$. 

Throughout the simulations we apply an equation of state with a constant adiabatic index of $ 4/3 $, as appropriate for a radiation dominated gas. We neglect gravity, as the gravitational dynamical times are longer than the typical interaction timescales.

We examine three setups: \A) $ v_{max} = 0.2c $ (the typical maximal velocity found in \citealt{Hotokezaka2012}) and $t_{inj}=80$ ms, \B) $v_{max} = 0.2c $ and $t_{inj}=240$ ms, and \C) $ v_{max} = 0.4c $ and $t_{inj}=40$ ms (the canonical model of \citealt{Nagakura2014}). In all cases we run the simulation up to a point where the cocoon expands ballistically: $1.2$ s in simulations \A and \C, and $1.4$ s in simulation \B. The details of the various setups are listed in table 1. The duration over which the jet is injected before it breaks out of the wind ($ t_{\rm{bo}}- t_{\rm{inj}}$) varies between 0.1 and 0.2 sec.

\begin{table}
\setlength{\tabcolsep}{18.5pt}
\centering
\begin{tabular}{ | l | c  c  c | }
    \hline
    Model & $ A $ & $ B $ & $ C $ \\ \hline
    $ M_{ej} [\msun] $ & $ 0.01 $ & $ 0.01 $ & $ 0.01 $ \\
    $ v_{max}/c $ & $ 0.2 $ & $ 0.2 $ & $ 0.4 $ \\
    $ v_{avg}/c $ & $ 0.12 $ & $ 0.12 $ & $ 0.23 $ \\
    $ t_{\rm{inj}}~[\rm{s}] $ & $ 0.08 $ & $ 0.24 $ & $ 0.04 $ \\
    $ L_j~[10^{50}\rm{erg~s^{-1}}] $ & $ 4 $ & $ 4 $ & $ 4 $ \\
    $ \theta_j $ & $ 10^\circ $ & $ 10^\circ $ & $ 10^\circ $ \\
    $ (t_{\rm{bo}}- t_{\rm{inj}})~[\rm{s}] $ & $ 0.1 $ & $ 0.2 $ & $ 0.12 $ \\
    $ R_{\rm{bo}}~[10^9\rm{cm}] $ & $ 1.4 $ & $ 3 $ & $ 2.5 $ \\
    $ E_c~[10^{49}\rm{erg}] $ & $ 3 $ & $ 4 $ & $ 2 $ \\
    $ t_{\rm{end}}~[\rm{s}] $ & $ 1.2 $ & $ 1.4 $ & $ 1.2 $ \\ \hline
\end{tabular}
\caption{The simulations configurations. $ M_{ej} $ is the ejecta mass, $ v_{max} $ is the ejecta front velocity and $ v_{avg} $ is its mass-weighted velocity, $ t_{\rm{inj}} $ is the time in which jet injection starts, $ L_j $ is the total jet luminosity, $ \theta_j $ is the jet launching opening angle, $ (t_{\rm{bo}}- t_{\rm{inj}}) $ is the time it takes the jet to break out of the ejecta since injection starts, $ R_{\rm{bo}} $ is the breakout radius, $ E_c $ is the cocoon energy, and $ t_{\rm{end}} $ is the simulation termination time. }\label{tab_models_comparison}
\end{table}

We carry out the numerical simulations using the special relativistic hydrodynamical module of the code PLUTO, version 4.0 \citep{Mignone2007}. We use an HLL Riemann solver and a third order Runge Kutta time stepping. Our 3D model employs a Cartesian grid. The $z$-axis extends from $z_{base}$ up to $ 4 \times 10^{10}\rm{cm} $ while the $x$ and $y$ axes extend from $ -4 \times 10^{10}\rm{cm} $ to $ 4 \times 10^{10}\rm{cm} $. In order to obtain a high enough resolution along the jet axis the Cartesian grid has two patches on each axis. An inner, high resolution patch in $ \pm 10^8~\rm{cm} $ for $ x $ and $y$ and from $ z_{base} $ to $ z = 2 \times 10^9~\rm{cm} $ on the $ z $ axis and, an outer logarithmic grid. In simulation \A the cell sizes in the inner region is $ 1.3 \times 10^6~\rm{cm} $, $ 1.3 \times 10^6~\rm{cm} $ and $ 3.3 \times10^6~\rm{cm} $ in $x$, $y$ and $z$ respectively, and the total number of cells is $ 800 \times 800 \times 1325 $. In simulations \B and \C the cells are larger by a factor of $4/3$ and the number of cells is lower by that factor in each axis\footnote{See Appendix A for a discussion of the numerical convergence.}.
We also carried out 2D simulations. However, due to their questionable credibility that we discuss briefly in $ \S $ \ref{sec_2d}, we do not elaborate on their numerical setup and results, and focus on the 3D simulations.

To check that the results do not depend strongly on the grid resolution, we checked different grid cell sizes and found that the energy roughly converges in these resolutions (see appendix \ref{sec_convergence}). Additionally, we checked the dependency on the Riemann solver as we have reconstructed simulation \A with a tvdlf solver and verified that the cocoon properties are similar at $ (t_{\rm{bo}} $.

\section{Hydrodynamic Results}
\label{sec_hydro}

\begin{figure}
	\includegraphics[width=\columnwidth]{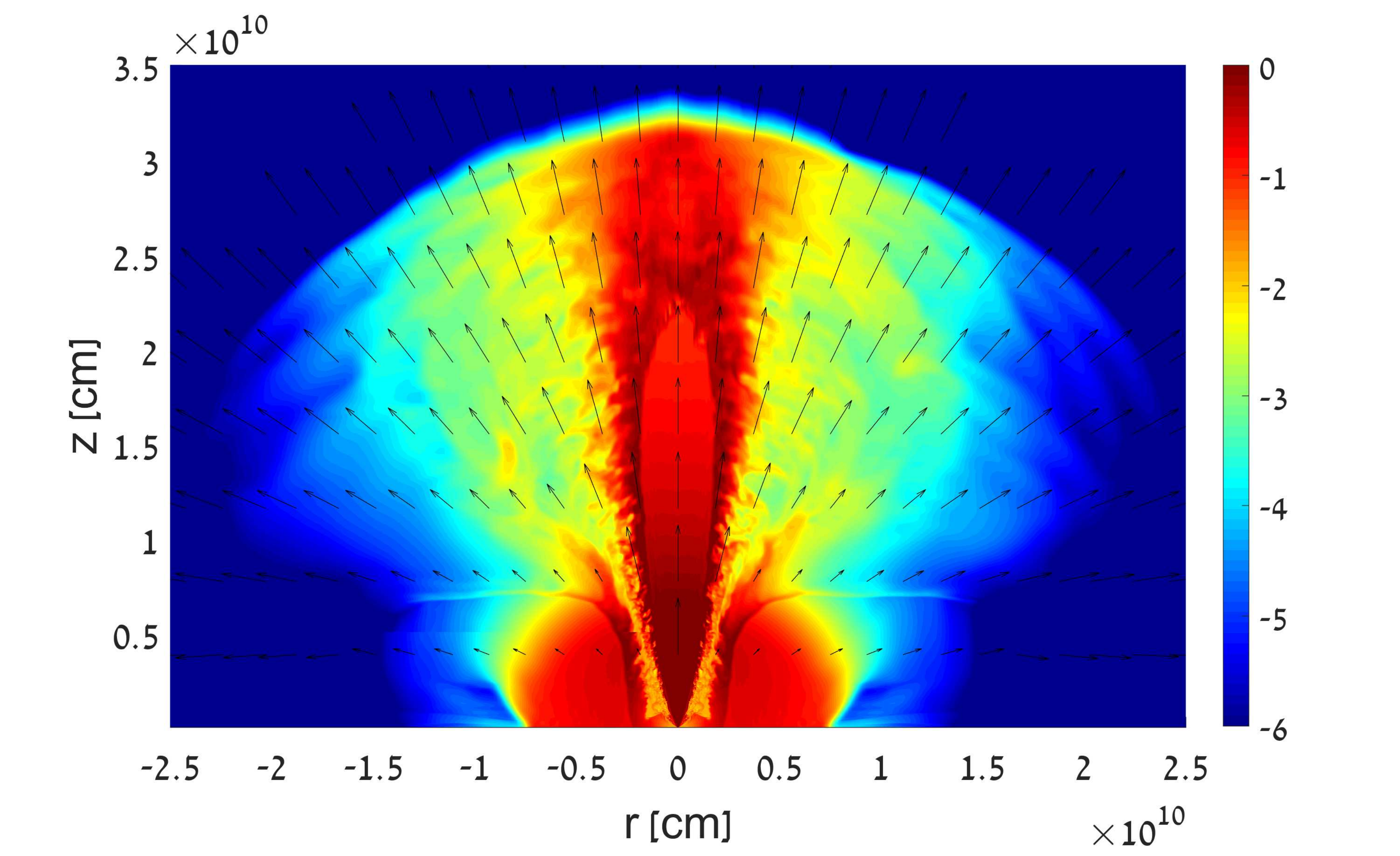}
	\caption{An energy density map of simulation \A in at the final snapshot, ($t = 1.2~\rm{s}$. The color scheme is logarithmic and it represents the energy density, excluding rest-mass, in the lab frame in units of $gr \cdot c^2$, namely $\Gamma(\Gamma-1)\rho+(4p\Gamma^2-p)/c^2$. The arrows mark the direction of the velocity and their length is proportional to the velocity size.
	(An animation is available in the online journal.)}
	\label{fig_3D_map}
\end{figure}

The general properties of all three simulations are similar. Figure \ref{fig_3D_map} shows a map of the logarithmic energy density of simulation \A at the end of the computations. 
The figure is taken at $t=1.2$ s, long after the jet broke out from the ejecta (at $ t = 0.18~\rm{s} $) and the cocoon material motion becomes ballistic. 
The following components can be identified: the jet, that is injected near the origin (in red), the ejecta that expands spherically from the origin (also in red), the outer cocoon (in cyan) and the inner cocoon that engulfs the jet (in yellow).

\begin{figure}
\includegraphics[width=\columnwidth]{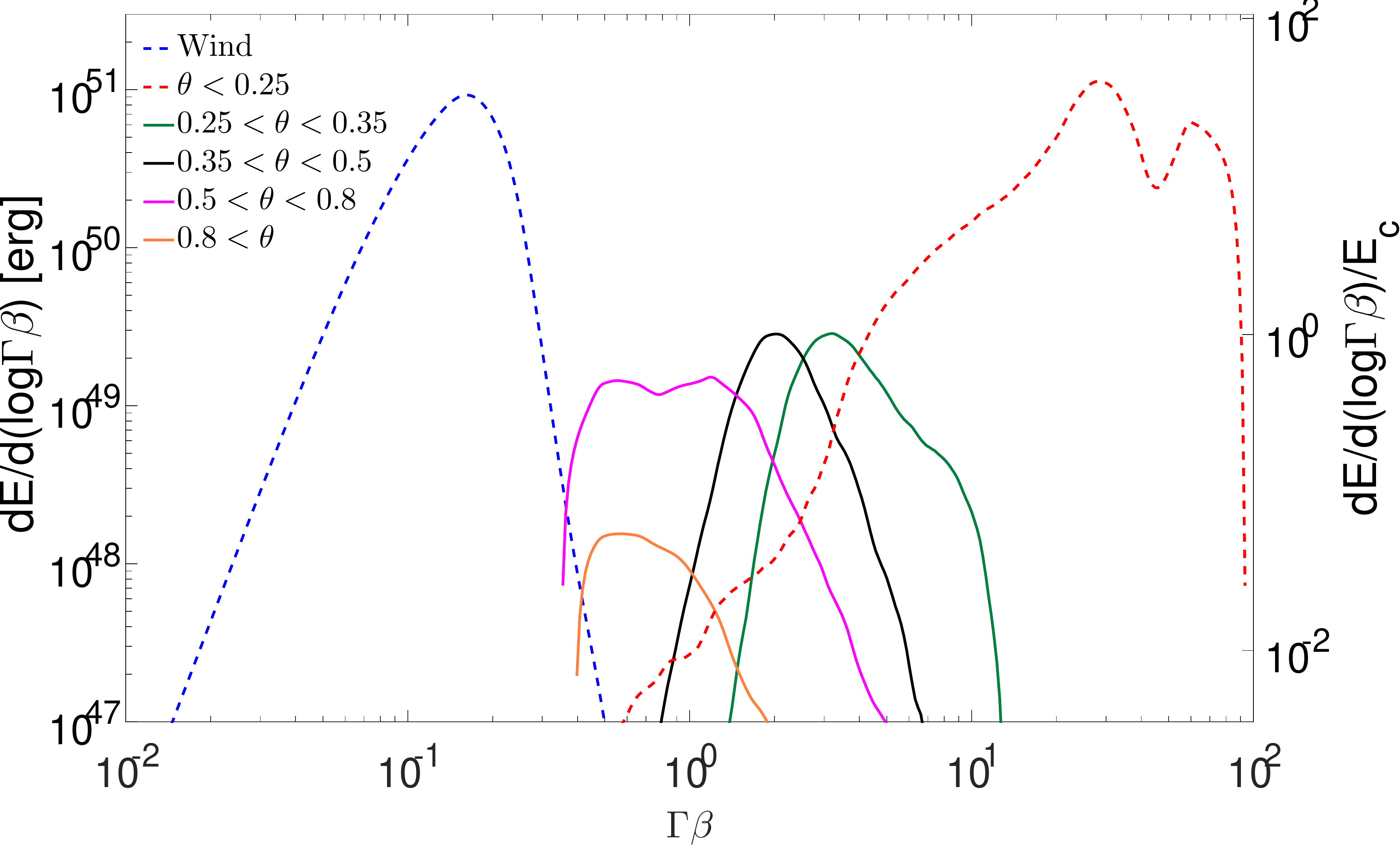}
\caption{ The distribution of the outflow (two sided) energy of various components per a logarithmic scale of $ \Gamma\beta $ for simulation \A at the end of the simulation ($t = 1.2~\rm{s} $). The distribution is given in cgs units (left) and in units of the cocoon's total energy $ E_c $ (right). The components that are not related to the cocoon are represented in dashed curves. We include all the material that is within an opening angle of 0.25 rad in the jet ({\it dashed red}), and all the material that is out of this opening angle but at radius that is smaller than that of the spherical ejecta front as part of the wind. The rest of the material is considered as part of the cocoon ({\it solid lines}) and is divided to various components according to the angle with respect to the jet axis. The figure shows that the energy distribution of the cocoon is roughly uniform in the four-velocity log space, where material with higher velocity dominates the outflow at smaller angles.}
\label{fig_3D}
\end{figure}

Figure \ref{fig_3D} depicts the corresponding energy distribution as a function of $ \Gamma\beta $.
The material can be divided to either one of three regions: (i) The material that moves at $ v \gtrsim v_{max} $ and has already broken out of the ejecta. This material generates the cocoon emission. By the end of the simulation it has reached its terminal velocity and its evolution at later times can be extrapolated from the last snapshots of the simulations. (ii) The slower material, which is still locked in the ejecta. This component is slower than the ejecta's front, so it will stay locked and it will continue to evolve with the isotropic ejecta.
(iii) Jet material that is still buried in the ejecta, and has yet to accelerate to its terminal velocity. When it reaches this stage, it becomes a part of the jet's bulk, so its contribution applies only to the jet's opening angle without affecting the cocoon. 
	
The total energy in the cocoon is $ E_c \approx L_j(t_{bo}-t_{inj}-R_{bo}/c) $, where $ t_{bo}-t_{inj} $ is the time over which the jet is injected before it breaks out of the ejecta and $ R_{bo} $ is the breakout radius \citep{Lazzati2005,Bromberg2011}. In the three simulations we carried out $E_c$ is in the range $2-4 \times 10^{49}$ erg. 
The energy distribution, depicted in Figure \ref{fig_3D}, reflects the different components of the system and their observational significance. The isotropic wind ejecta will ultimately produce the late time macronova signal. At $\theta<0.25$ rad the outflow is dominated by the jet that produces the SGRB. This outflow is ultra-relativistic and its emission can be seen only by on-axis observers. 
The cocoon component is spread over a wide solid angle and it spans a wide range of $\Gamma \beta$ values. Its opening angle is about $45^\circ$ and its energy distribution with the four-velocity 
spreads roughly uniformly in log space between the wind maximal velocity and $\Gamma\beta \approx 5$. About $10\%$ of the cocoon energy is carried by Newtonian material at angle $>45^\circ$.

\subsection{2D simulations}	
\label{sec_2d}
We have also carried out 2D simulations of the same setups. These simulations enable us to use a higher resolution and thus they reach a better numerical accuracy. However, we find that for the specific purpose of finding the cocoon emission, the imposed unphysical axial symmetry render their results unreliable. Specifically, in all 2D jet simulations there is a plug of heavy ejecta material that accumulates on top of the jet head \citep[e.g.,][]{Lazzati2009,Mizuta:2013}. The plug is not vacated aside to the cocoon during the entire propagation and in our simulations it remains in front of the jet also after the breakout throughout the entire duration of the simulation. 
The jet-plug interaction has two effects. First, as the interaction continues long after the jet breaks out, it increases the cocoon energy significantly. Second, jet material that is deflected by the plug has a relatively high Lorentz factor. Thus, the interaction with the plug ultimately results in a significant amount of material with a Lorentz factor $\Gamma \sim 10$ that expands at a wide opening angle. In the 3D simulations the symmetry is broken and there is no significant plug (see also \citealt{Zhang2003,harrison17}). As a result the wide angle $\Gamma \sim 10$ material is also absent. This material, which seems to be a numerical artifact of the symmetry imposed in 2D, has a profound effect on the emission, especially during the cocoon afterglow phase (see $ \S $ \ref{sec:afterglow}). Hence, we use here only the results of the 3D simulations. We verify, though, that regions that are unaffected by the plug are similar in 2D and 3D simulations. We leave a detailed discussion of this and other 2D artifacts to a future work (Gottlieb et al. 2017, in preparation).

\section{Emission} \label{sec_emission}
\subsection{Method} \label{sec_calculation}

We calculate the emission by post processing the final snapshots of the hydrodynamical simulations. The cocoon emission has three components (NP17): diffusion of the internal energy deposited by shocks (cooling emission), radioactive heating (macronova) and interaction with external medium (afterglow). Here we calculate only the first two and discuss the afterglow separately ($ \S $ \ref{sec:afterglow}), as we do not expect it to have a significant contribution. Since we are interested only in the cocoon emission, we calculate the contribution only from material that is ahead of the spherical wind at the time that our hydrodynamic simulations end. Thus, our calculated light curves do no include the standard macronova emission from the spherical ejecta. 

Given that the cocoon material expands ballistically at the end of the hydrodynamical simulation we can extrapolate the hydrodynamic variables of each cocoon fluid element to any time as long as the radiation is still trapped. For each time step we calculate first the ``radial" optical depth from every radius to infinity along a radial path. We do that for different angles and we determine the trapping radius, $r_t(\theta,\phi)$, where $\tau(r_t(\theta,\phi))=c/v$. Above this radius photons diffuse freely to the observer at infinity while below this radius they are trapped. This is an approximation since the outflow is not spherically symmetric. In a similar manner we calculate for each angle {the photospheric radius} $r_{ph} (\theta,\phi)$ for which $\tau(r_{ph}(\theta,\phi))=1$. 

We calculate separately the luminosity of the cooling emission and the macronova radioactive heating. The cooling emission in the fluid frame is the diffusing rest frame energy flux at $r_t$ of the radiation that was carried by the outflow from the last hydrodynamical snapshot. We can find this flux since the radiation at $r_t$ is trapped up to time $t$, carried by the outflow and cools adiabatically between the last hydrodynamical snapshot and the time $t$.

{The macronova emission arises from the radioactive heating generated by material above $r_t$.}
To calculate the macronova luminosity {we follow the approximations used by \cite{Grossman2013}. We} find, along each radial trajectory, the mass at 
$r>r_t$ and multiply it by the radioactive heating rate. The instantaneous rest frame energy deposition rate is: $\dot{\epsilon} \Big (t'/1\rm{s}\Big)^{-1.3} m(r>r_t)$ where $t'$ is comoving time. The value of $\dot{\epsilon}$ can vary by a factor of few depending on the outflow composition and time. We consider two values. The first, $\dot{\epsilon}=\dot{\epsilon}_0$, where $\dot{\epsilon}_0 = 10^{10}~\rm{erg~gr^{-1}~s^{-1}} $ \citep{Freiburghaus1999,Metzger2010,Korobkin2012,Hotokezaka2017}, which is typical for dynamic ejecta that is dominated by elements with mass number $A>130$. The second, $\dot{\epsilon}=2\dot{\epsilon}_0$, is expected at the relevant times ($\sim 10^4$ s) for a neutrino driven wind dominated by lighter ($ A < 130 $) elements \citep{Perego2014,Martin2015}.

We estimate the rest frame temperature by taking { for each direction, $\theta$ and $\phi$,} the rest frame luminosity (the sum of cooling and macronova emission) and finding the energy density at the radius $r_{ph}(\theta,\phi)$ assuming that the radiation is in a thermal equilibrium and that the local radiation spectrum is a blackbody at this point. 
We assume now that this black body radiation is emitted isotropically at the matter's rest frame at this point.
Having the rest frame luminosity and spectrum along each angle at every time in the explosion frame, we integrate the contribution from material at all angles for observers at different viewing angles at different observer times, by properly accounting for the Lorentz boost and the light travel time.

The opacity, $ \kappa $, of r-process elements play a decisive role. At the same time it is quite uncertain. Two representative values that are discussed in the literature are $ \kappa = 10~\rm{cm^2~gr^{-1}} $ \citep{Kasen2013,Tanaka2013} for material containing Lanthanides and Actinides, and $ \kappa = 1~\rm{cm^2~gr^{-1}} $ for lighter heavy elements of mass numbers $ A < 130 $ \citep{Martin2015}. The latter is particularly relevant in neutrino driven winds that may dominate at high latitudes \citep{Perego2014}.
We compare two numerical calculations with these two representative opacities and the two corresponding energy injection rates. We estimate that the unknown opacity and the assumption of a blackbody spectrum are the main source of uncertainty in our calculations.
The values of $\kappa$ that we consider do not necessarily cover the full range of possible values and the spectrum may deviate significantly from a blackbody. For instance, the cocoon's cooling emission is found to be dominated by the Rayleigh-Jeans tail of a very hot plasma ($ T \gg 10^4~\rm{K} $). The corresponding opacities and spectra for these temperatures are yet to be explored.
	
\subsection{Numerical Light Curves}
\label{sec_photospheric_results}

\begin{figure*}
	\centering
	\includegraphics[scale=0.245]{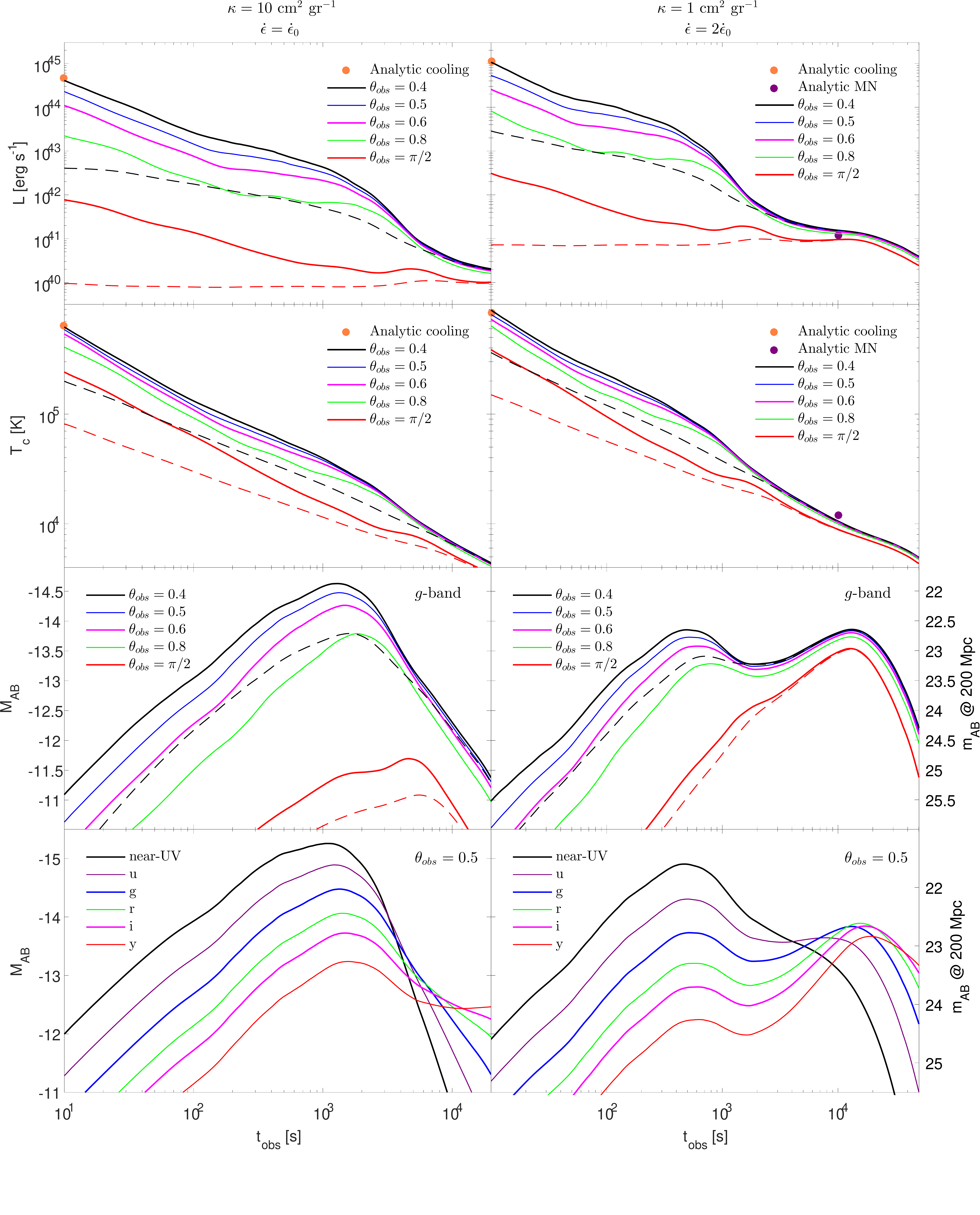}
	\caption{The results of simulation {\it A}. Bolometric luminosity {\it (top row)}, color temperature {\it (second row)}, $ g $-band {\it (third row)} and multi-wavelength {\it (bottom row)} AB magnitudes (absolute on the left and apparent for observer at 200Mpc on the right). The first three rows are for observers at different angles with respect to the jet axis, while the bottom row is for an observer at $\theta_{obs}=0.5$ rad. In the left column the opacity is $\kappa=10 ~\rm{cm^2~gr^{-1}}$ and the energy injection rate coefficient is $\dot{\epsilon}=\dot{\epsilon}_0$. In the right column the opacity is $\kappa=1 ~\rm{cm^2~gr^{-1}}$ and $\dot{\epsilon}=2\dot{\epsilon}_0$. The solid lines mark the total luminosity (the sum of the cooling emission and the macronova), while the dashed curves depict the luminosity and magnitude arising from the macronova emission alone. The analytic estimates are taken from NP17 using the cocoon properties given in table 1 and $f_\Gamma$ as measured in the simulations.}\label{fig_emission2}
\end{figure*}

Figure \ref{fig_emission2} depicts the light curve resulting from simulation {\it A}. Shown are the bolometric luminosity, color temperature and $g$-band magnitude for observers at different angles with respect to the jet axis, as well as light curves in several IR/opt/UV bands seen by an observer at $\theta_{obs}=0.5$ rad. These are given for the two representative opacities and the two corresponding energy injection rates, namely ejecta with $\kappa=10 ~\rm{cm^2~gr^{-1}}$ and $\dot{\epsilon}=\dot{\epsilon}_0$ and ejecta with $\kappa=1 ~\rm{cm^2~gr^{-1}}$ and $\dot{\epsilon}=2\dot{\epsilon}_0$. It also shows the analytic prediction of NP17 for the cooling emission at $t_{obs} = 10$ s when the emission is dominated by material with $\Gamma \approx 3$ (at later time, during the mildly relativistic phase, the analytic model is less accurate), and for the macronova emission from the shocked ejecta. Both predictions agree very well with the numerical results. $E_c$, $R_{bo}$ and $f_\Gamma$ for the analytic model are taken from the numerical simulation, where $ f_\Gamma $ is the fraction of the cocoon energy stored in a logarithmic scale of $ \Gamma\beta $. As discussed in $ \S $ \ref{sec_hydro}, the cocoon energy is distributed uniformly in log space over two logarithmic scales, implying that $ f_\Gamma $ is roughly constant, and that it equals $ \sim 0.5 $. The properties of the cocoon that we find are very different than the spherical outflow with $\Gamma=10$ assumed by \cite{Lazzati2016} and therefore their light curve predictions are also different than our findings here.

During the first minute after the merger the emission towards an observer at $\theta_{obs}<0.8$ rad is very bright, $\sim 10^{45}~\rm{erg~s^{-1}}$. However, this luminosity is not accessible to the observer since it is radiated in the extreme UV, at $T_c \sim 10^5-10^6$ K. During the first $10^3$ s the emission is dominated by material with $\Gamma \sim 2-3$ and the luminosity and temperature falls roughly as $L \propto t_{obs}^{-1.2}$ and $T_c \propto t_{obs}^{-0.7}$. As a result the luminosity of the UV/opt/IR, which are at the Rayleigh-Jeans part of the spectrum, rises. Around $10^3-10^4$ s there is a transition to the Newtonian phase (i.e., the velocity at $r_t$ becomes non-relativistic).
During the transition the luminosity drops more rapidly and the high latitude emission approaches that of the low latitude. This evolution leads to a monochromatic blue peak at the beginning of the transition, $t_{obs} \sim 10^3$, which is dominated by the cooling emission of material with $\Gamma \approx 1.4$. The luminosity and timing of the peak depends on the opacity. A larger opacity leads to a brighter peak at a later time. During the Newtonian phase the radioactive heating becomes the main luminosity source, giving rise to a second,
almost isotropic, red and chromatic peak on time scales of hours to a day. Here a larger opacity leads to a fainter, redder and longer duration peak (similarly to the main macronova event).

\begin{figure*}
	\includegraphics[scale=0.25]{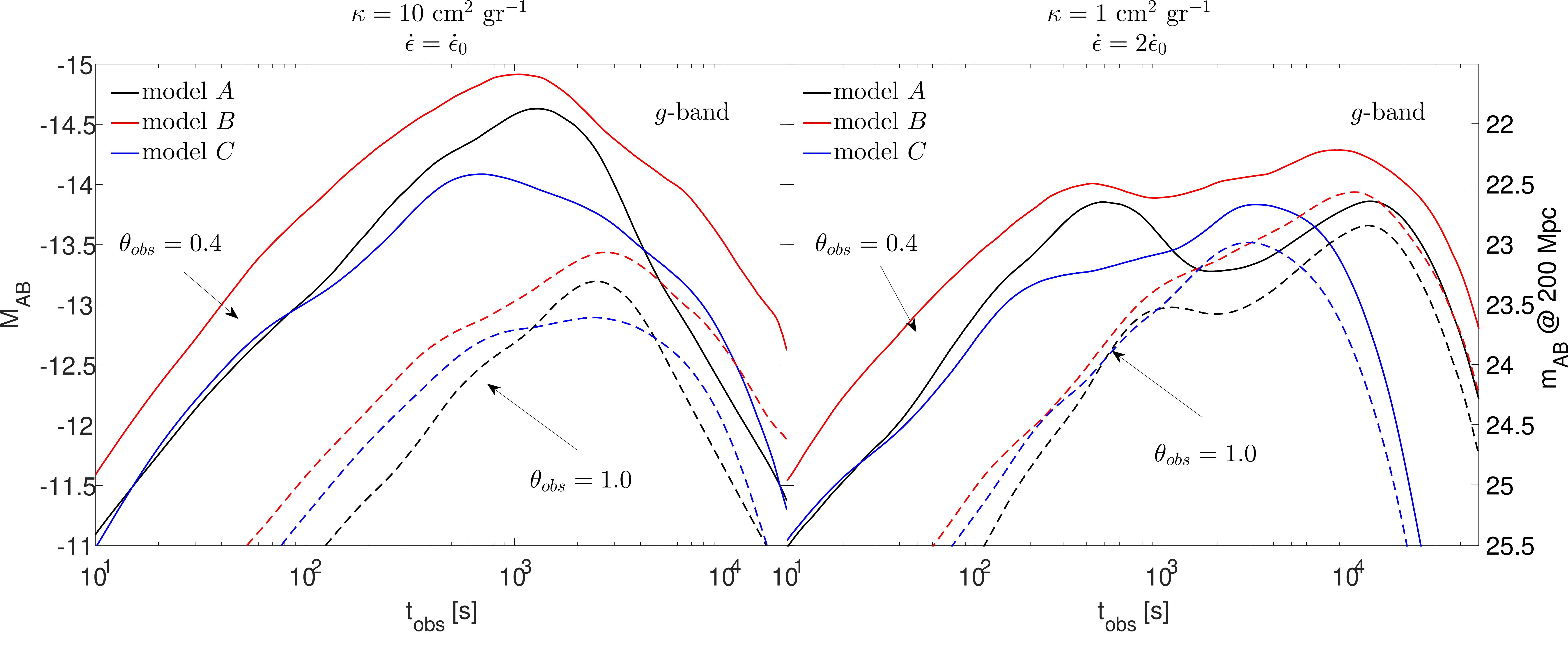}
	\caption
	{$g$-band light curves of the three simulations (\A, \B and \C) as seen by an observer at $\theta_{obs}=0.4$ rad ({\it solid}) and $\theta_{obs}=1$ rad ({\it dashed}). On the left the opacity is $\kappa=10 ~\rm{cm^2~gr^{-1}}$ and the energy injection rate coefficient is $\dot{\epsilon}=\dot{\epsilon}_0$. On the right $\kappa=1 ~\rm{cm^2~gr^{-1}}$ and $\dot{\epsilon}=2\dot{\epsilon}_0$.}
	\label{fig:ModelsComparison}
\end{figure*}

Simulations \B and \C show a similar behaviour to the one observed in simulation \A, although there are some differences in the details. A comparison of the $g$-band light curves of the three simulations is shown in figure \ref{fig:ModelsComparison}. Simulation {\it B}, which has the most energetic cocoon and the largest breakout radius, produces the brightest emission. For the simulations' parameters that we use here all models produce a rather bright emission with absolute magnitudes that ranges between -14 and -15 on time scales of $10^3$-$10^4$ s. In order to extrapolate our results to a range of cocoon parameters we can use the analytic formula from NP17 and approximate the peak of the cooling emission as being observed at the end of the relativistic regime. Then using equations $28-30$ in NP17 (keeping $ f_\Gamma $ and $\Gamma$ constant) and approximating the spectrum as a blackbody we obtain that in the Rayleigh-Jeans regime $L_\nu \propto E_c^{3/4}R_{bo}^{1/4}\kappa^{1/2}$. This scaling is in a rough agreement with the results of the three simulations. The properties of the second, radioactive, peak can be estimated analytically from equations $13$, $35$ \& $36$ of NP17 that are also at agreement with our numerical results.

\subsection{Cocoon afterglow}\label{sec:afterglow}

Additional emission arises from the interaction of the expanding cocoon material with the circum-merger medium. This emission depends, of course, on the medium density. It is also extremely sensitive to the outflow's Lorentz factor. Material with $\Gamma \approx 10$ produces a signal that peaks on a time scale of hours in the X-ray and Optical. In all our 3D simulations\footnote{In 2D simulations a significant fraction of the cocoon energy is carried over a wide angle by $\Gamma \approx 10$ material. This seems to be an artificial artifact of the imposed axial symmetry. This is the main reason that we avoid using 2D simulations (see $ \S $ \ref{sec_2d}). } only a negligible amount of energy is carried by such material ($<1\%$ of the cocoon energy) and even this small amount of energy is confined to a narrow angle ($<15^\circ$). Therefore we expect that the cocoon afterglow will not produce an observable X-ray or optical signals, at least for the configurations that we have examined.

The fastest material that we see in our simulations may produce an observable radio signal. About $10^{49}$ erg are carried by material with $\Gamma \approx 3$ over an opening angle of $20^\circ$. If the external density is not too low ($\sim 0.1~{\rm cm^{-3}}$) then, assuming typical afterglow parameters, this material produces a radio signal on a time scale
of weeks at a level of $\sim 0.1$ mJy over an opening angle of about $45^\circ$ at a distance of 200 Mpc \citep{Nakar2011,Piran2012,Hotokezaka2015}. However, in general we expect this cocoon radio signal to be comparable or lower than the off-axis GRB radio afterglow. The off-axis afterglow peaks once the afterglow blast wave decelerates to a Lorentz factor that is similar to the cocoon Lorentz factor. At this stage the two signals depend on their energies. Since the GRB jet energy is expected be at least as high as that of the cocoon, its off-axis radio emission is comparable or brighter.

\section{Summary}
\label{sec_summary}

Double NS mergers are a major target for GW detectors and the detection of their EM counterparts are, therefore, extremely important. 
It is generally believed that SGRBs are associated with these mergers and observations and numerical simulations suggest that such mergers are surrounded by a quasi-spherical ejecta. 
The propagation of the SGRB jet within the ejecta surrounding the merger produces a cocoon that carries out a significant amount of energy. The cooling emission of this cocoon and its radioactive macronova signature are short lived UV-optical signals that are potentially strong enough to be observed up to the advanced LIGO detection horizon. 
		
We present here the first numerical simulations of a cocoon emission light curve from a SGRB. We have carried out 3D relativistic hydrodynamic simulations of the jet propagation within the merger's ejecta. For the merger's outflow we follow \citet{Nagakura2014} who based their double NS merger simulation of \citet{Hotokezaka2012}. We carry out the simulations 
until the motion of the cocoon material becomes ballistic. We then use our hydrodynamic results to calculate the cocoon emission, considering two processes: the cooling emission and the radioactive macronova.

The initial bolometric luminosity is very high, reaching more than $10^{44}~\rm{erg~s^{-1}}$, but it peaks in the extreme UV. During the first $\sim 10^3$ s 
the bolometric luminosity decreases. However, due to the drop in the temperature 
the signals in the UV/optical bands rise while the ejecta expands (similarly to the early time cooling emission from supernovae). The cooling emission peaks 
around a thousand seconds after the merger and, in the setups we considered, it reaches -14$^{th}$ to -15$^{th}$ absolute magnitudes in the g-band over a relatively wide opening angle. 
The cocoon's macronova is more important for lower opacities ($\kappa \sim 1~{\rm cm^2~gr^{-1}}$). It peaks at a slightly later time ($\sim 10^4$ s) and it has a somewhat weaker signal (about 14$^{th}$ $g$-magnitude). This emission is isotropic and of longer duration in comparison to the cooling emission.

The signal that we find depends, first, on the parameters of the cocoon, which in turn depend on the properties of the ejecta and the jet, and on the delay (if any) between the merger and the launch of the jet. In general, the longer it takes the jet to break out of the cocoon, the more energetic is the cocoon and the larger is the breakout radius. This results in a brighter signal. In this paper we have considered several plausible setups, but the uncertainties of the jet and ejecta properties as well as the delay are large and the actual signal may be significantly brighter or fainter than what we find here. We note that we considered setups where the jet breaks out of the ejecta within 0.1-0.2 s after being launched. If, as suggested by the results of \cite{Moharana2017}, the typical breakout time is about $ 0.4$ s, then the cocoon signal may be brighter than the one we find here.
{Another source of uncertainty is the poorly constrained opacity.
The signal we find is composed of two components, an early cooling peak and a later macronova signal. For higher opacity the cooling emission peaks later and is brighter, while a lower opacity magnifies the radioactive macronova signal. For the cases we consider the two peaks are comparable for $\kappa = 1-10~{\rm cm^2~gr^{-1}} $, and any opacity within the range of $ 0.1-100~{\rm cm^2~gr^{-1}} $ would yield a signal as bright or brighter.}

Advanced LIGO is designed to detect double neutron star mergers out to $ 200~\rm{Mpc} $ \citep{LIGOScientificCollaboration2010}. The signal at the levels that we find here has an apparent magnitude of $\approx 22$ at that distance. It will be easily detected by the Large Synoptic Survey Telescope (LSST; \citealt{LSSTScienceCollaboration2009}). Near future less sensitive large field-of-view surveys, such as the Zwicky Transient Facility (ZTF; \citealt{Bellm2014}), may also be able to detect this signal if it is slightly brighter or if the merger takes place slightly closer. We note that due to the very blue signal of the cooling emission a rapid search in the UV may be very well paid out. At a distance of 200 Mpc the signal that we find will be detectable by the planned UV satellite ULTRASAT \citep{sagiv2014} that, with its broad field of view, seems to be an ideal telescope to search for this EM counterpart following a GW detection.

Two other candidate short lived blue signals have been proposed. A signal arising from the neutrino driven wind \citep{Grossman2013,Prerego2014,Kasen2015} and a signal arising from a neutron rich outflow that may precede this wind \citep{Metzger2015}. While less certain (both signals can be suppressed by a contamination of Lanthanides that will increase the opacity) and slightly dimmer than the cocoon signals that we find here, they have similar characteristics. 
The existence of these blue and rather strong but short lived signals suggests that the best
strategy to detect an EM counterpart to an advanced LIGO merger trigger would be a 
very quick search using wide field telescopes in the near-UV and blue bands followed by a deep search in the IR in candidate spots, discovered in the optical and near-UV \citep{Grossman2013}. 

\section*{Acknowledgements}
This research was supported by the I-Core center of excellence of the CHE-ISF. TP was partially supported by an advanced ERC grant TReX. OG and EN were partially supported by an ERC starting grant (GRB/SN) and an ISF grant (1277/13) and an ISA grant.

\bibliographystyle{mnras} 
\bibliography{SGRB_mnras}

\appendix

\section{Numerical convergence}
\label{sec_convergence}

\begin{figure}
	\centering
		\includegraphics[scale=0.19]{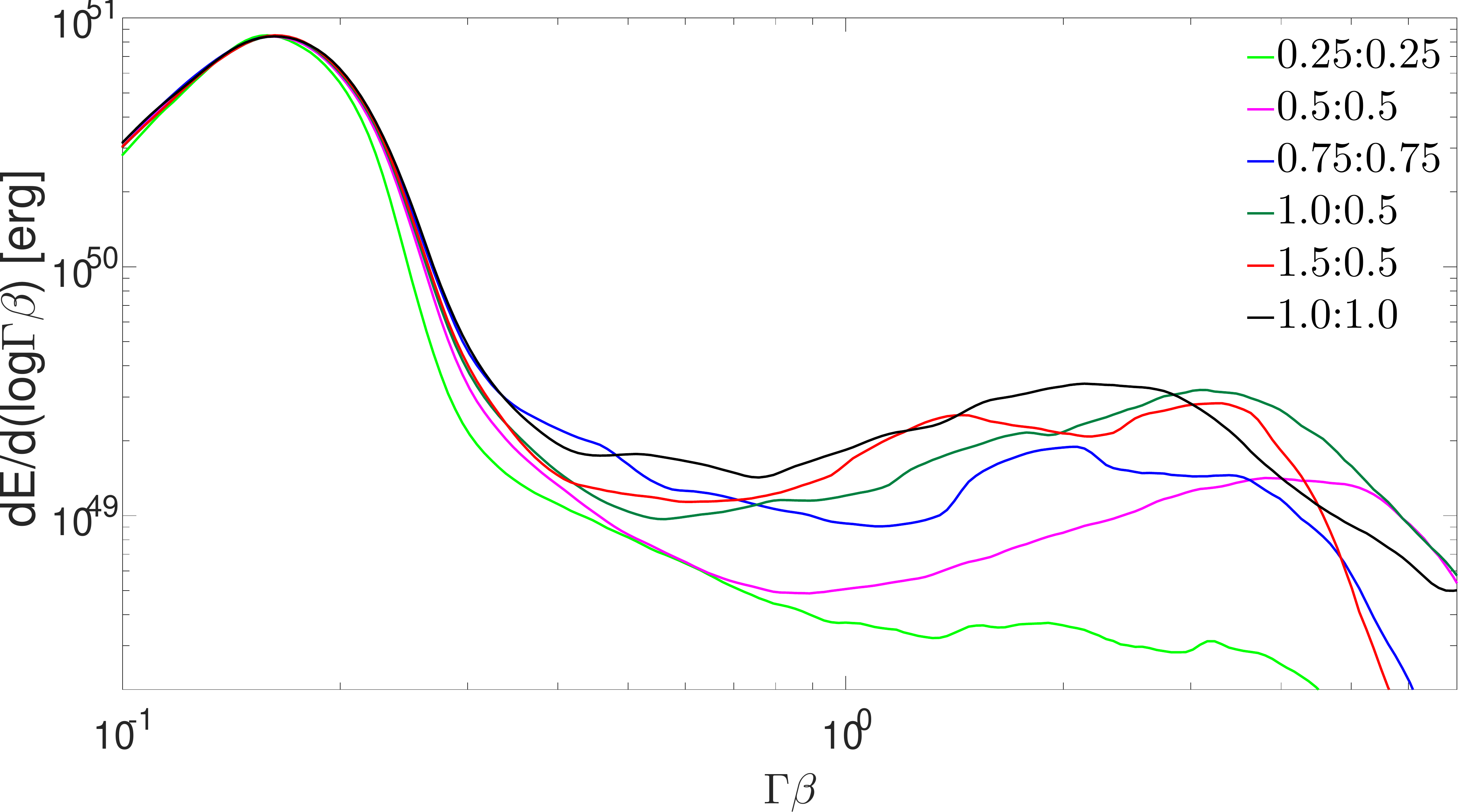}
	\caption{Comparison between simulations with different resolutions of the energy distribution as a function of $ \Gamma\beta $ at $ t = 0.65 $s. All simulations run the same setup of simulation \A but with different resolutions, which are marked in the legend. The first number marks the resolution of the inner patch, along the jet, while the second marks the resolution in the outer patch. For example 1.5:0.5 is a simulation where the length of the cells in the inner patch is shorter by a factor of 1.5 while in the length of the outer cells is larger by a factor of 2.}
	\label{fig_convergenceDist}
\end{figure}

\begin{figure}
	\centering
	\includegraphics[scale=0.3]{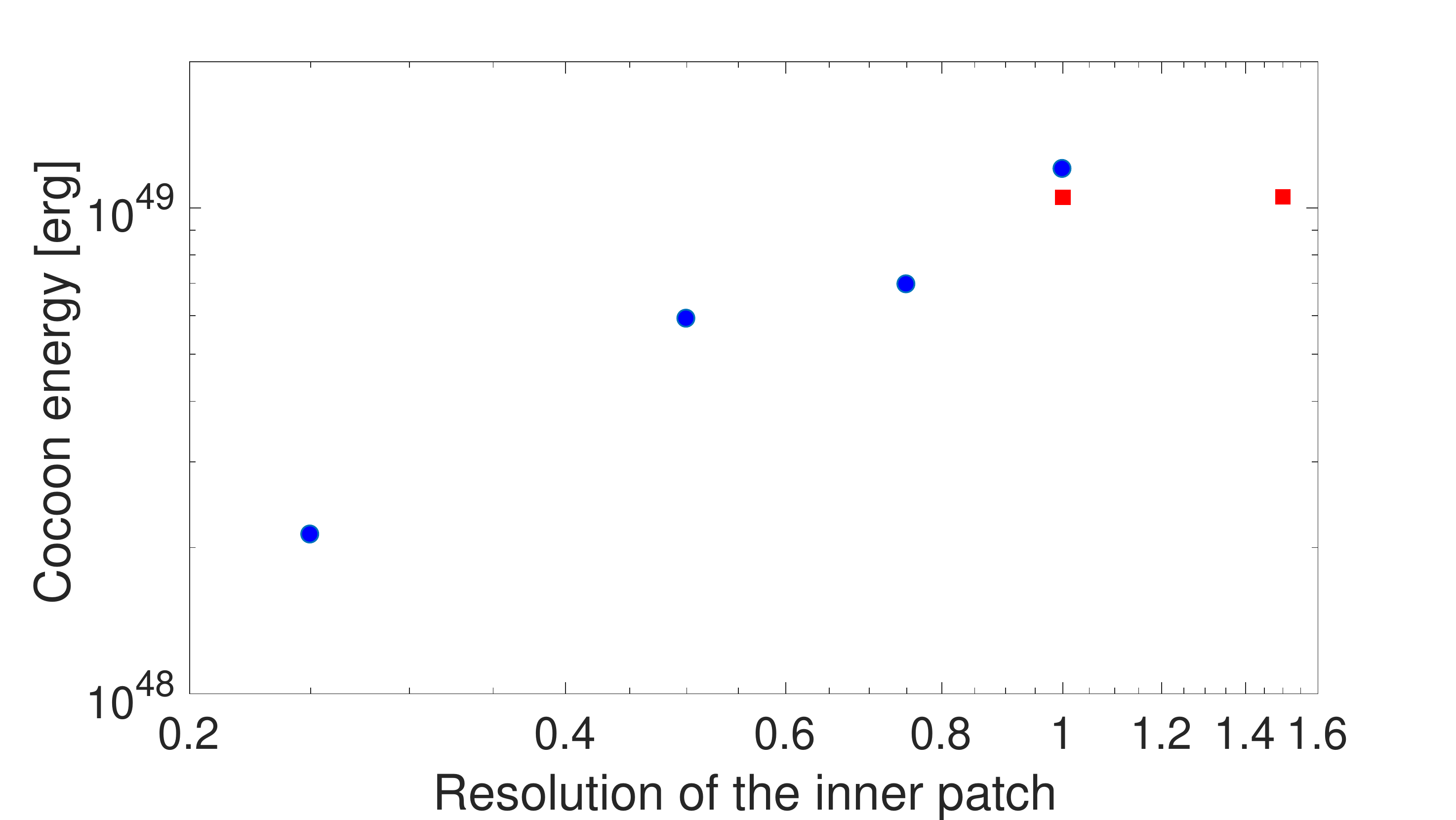}
	\caption{The two-sided cocoon energy in simulation \A as a function of the resolution in the inner patch (along the jet), normalized to the cell length of our canonical resolution (i.e., 0.25 means that the length of each cell in the inner patch is longer by a factor of 4 compared to our canonical resolution). The blue circles represent simulations where the ratio of the cells' length in the outer patch (compared to the canonical resolution) is similar to that of the inner patch. The red squares represent simulations where the resolution in the outer patch is 0.5 of the one in the canonical simulation. The energy is taken at $ t = 0.65 $s, after the cocoon energy stop evolving in time, and it contains material at $ \theta > 0.25$ rad and $0.5 < \beta\Gamma < 10 $ (i.e., excluding wind and jet material).}
	\label{fig_convergence}
\end{figure}

We have checked the convergence of our simulations by comparing the resolution used for simulation \A against lower resolutions: 0.25, 0.5 and 0.75 of the original resolution (i.e., the length of each of the cells' dimensions is larger by a factor of 4, 2 and 4/3 compared to simulation {\it A}). Figure \ref{fig_convergenceDist} shows a comparison between the various simulations of the energy distribution as a function of the four-velocity at $ \theta > 0.25$ rad. The cocoon material is seen at velocity $\gtrsim 0.5$c (at slower velocities the wind dominates). In all simulations the shapes of the distributions are rather similar, showing a relatively flat distribution or a moderate rise in the energy up to $\Gamma\beta \approx 4$. However, the normalization of the distributions (i.e., total energy) increases monotonically with the resolution. The increase in the total cocoon energy with the energy can be seen clearly in Figure \ref{fig_convergence}. 

The shape of the distribution is determined mostly by the mixing processes while the normalization is determined by the energy injection during the jet propagation into the cocoon (e.g., a jet with a shorter breakout time deposits less energy). Figure \ref{fig_convergenceDist} suggests that the mixing is not strongly affected by the resolution  while the total energy deposit does. To check this possibility we varied the relative resolution of the two patches in our grid (see section \ref{sec_numerical}). The inner patch contains the jet, and it therefore determines its propagation velocity, energy injection into the cocoon and mixing that takes place along the jet and in the head. The outer patch contains most of the cocoon and therefore determines a large part of the mixing, which takes place in the cocoon itself. We run two additional simulations where we used a reduced resolution in the outer patch (0.5 that of simulation \A), while keeping the inner patch resolution similar in one and increasing it by a factor of 1.5 in the other. Figures  \ref{fig_convergenceDist} and \ref{fig_convergence} show that indeed the total cocoon energies as well as the velocity distributions of these two simulations are very similar to that of simulation \A. 

Our conclusion is that simulation \A seems to have converged as far as the shape of the velocity distribution is concerned. The total energy also shows signs of convergence as the total energy does not change when the inner resolution is increased. However, we cannot be sure without having a simulation with a higher resolution over the entire grid, which is unfortunately impossible with our current computational resources (simulation \A takes 0.5M CPU hours). The tendency of the energy to increase with increasing resolution suggests that such simulations may yield somewhat more energetic cocoons that will result in brighter signals.
Finally, in simulations \B and \C we used a resolution that is 3/4 that of simulation \A. Therefore it is plausible that a higher resolution simulation of the same setups as \B and \C would result in a signal that is somewhat brighter. 
	
\bsp	
\label{lastpage}
\end{document}